\documentclass[10pt, oneside]{article} 
\usepackage{geometry}  
\usepackage{graphicx}
\usepackage{amssymb}
\usepackage{amsmath}
\usepackage{xcolor}
\usepackage{tikz}

\def\degree{{d}}

\def\firstEigenvalue{\Lambda_1}

\begin{document}

\begin{center}
{\Large Kuramoto model, phase synchronisation and the network's core}\\[3ex]
Ra\'ul J. Mondrag\'on\\
School of Electronic Engineering and Computer Science\\
Queen Mary University of London\\
London E1 4NS\\UK

\begin{minipage}{0.85\textwidth}
In this note we show that for the Kuramoto model defined in a simple  undirected graph it is possible to decide which nodes form the core of the network. The set of core--nodes is defined by its relevance to the phase synchronisation process.\\[3ex]
\end{minipage}\end{center}

The Kuramoto model is a nonlinear system of $N$ coupled oscillators with random frequencies $\omega_n$ and phases $\theta_n$ described by
\begin{equation}
\dot\theta_n = \omega_n+K\sum_{m=1}^N A_{nm}\sin(\theta_m-\theta_n),
\end{equation}
where $A_{nm}$  is the coupling term between phase $n$ and phase $m$ with strength $K$. If the strength $K$ is large enough the phase of, some or all, the oscillators tend to synchronise. 

If the coupling term is described by the simple--undirected graph ${\cal G}(E,V)$ then $A_{nm}$ are the entries of the adjacency matrix ${\bf A}$.  Restrepo~et.~al~\cite{restrepo2005onset} introduced the time--average approximation to estimate the critical value $K_c$ when the phases start synchronising. They describe the oscillators dynamics using local mean variable $r_n$ 
\begin{equation}
r_ne^{i\psi_n}=\sum_{m=1}^N A_{nm} \langle e^{i\theta_m}\rangle_t,\quad n=1,\ldots, N,
\label{eq:globalSynch}
\end{equation}
where $\langle \ldots \rangle_t$ is a time--average.
Under the assumptions that the minimum degree is large $\degree_{\rm min} \gg 1$ and that the solutions of $r_n$, $\psi_n$ are statistically independent of $\omega_n$ then 
\begin{equation}
r_n=\sum_{|\omega_m|\le Kr_m} A_{nm} \sqrt{1-\left(\frac{\omega_m}{Kr_m}\right)^2}.
\label{eq:restrepo}
\end{equation}

The critical value $K_c$ is estimated considering that the oscillators are not described by a particular sequence
of natural frequencies $\{\omega_i\}$ but rather by frequencies distributed according a unimodal probability density $g(\omega)$  with $g(-\omega)=g(\omega)$ then $K_c = {K_0}/{\firstEigenvalue}$, where $K_0 = 2/(\pi g(0))$ and  $\firstEigenvalue$ is the largest eigenvalue of the adjacency matrix ${\bf A}$. 

The purpose of this short communication is to define a network--core based on the high degree nodes  and their relevance to the synchronisation process. The approach is to  relate the connectivity of the high degree nodes with the eigenvalue $\firstEigenvalue$ and hence to the critical coupling value.

Assume that the nodes are ranked in decreasing order of their degree and that the top $m$ ranked nodes are the core of the network. An approximation to the eigenvector which corresponds to $\firstEigenvalue$ is constructed by considering the vector ${\bf v}(m)$ where the entry ${v}_n(m)$ is 1 if the node $n$ shares a link with a node in the core and zero otherwise. The entries ${u}_n(m)$ of ${\bf u}(m)={\bf A}{\bf v}(m)$ are the number of walks of length 1 which start in the node $n$ and end in a node belonging to the core. If $m=N$ then ${\bf u}_n(N)=\degree_n$  is the degree of node $n$.  
Based on the Rayleigh quotient an approximation $b(m)$ to  the eigenvalue is
\begin{equation}
b(m)= \frac{ {\bf u}(m)^{\tt T}{\bf A}{\bf u}(m) }{ {\bf u}(m)^{\tt T}{\bf u}(m)} 
\le \firstEigenvalue .
\end{equation}
The approximation $b(m)$ is a lower bound of $\firstEigenvalue$ as it is related to the bound based on walks of length two and three. If ${\bf v}=(1,\ldots, 1)$, then ${\bf u}^{\tt T}(N){\bf A}{\bf u}(N)={\bf v}^{\tt T}(N) {\bf A}^3 {\bf v}(N)= w_3$ is the total number of walks of length three and $ {\bf u}(N)^{\tt T}{\bf u}(N)={\bf v}^{\tt T}(N) {\bf A}^2 {\bf v}(N)=w_2$ is the number of walks of length 2, then it is known that $w_3/w_2 \le \firstEigenvalue$ (\cite{taubig2014matrix}, Th: 10  ). 

The nodes that form the core are the $m_c$ top ranked nodes where $\firstEigenvalue-b(m_c)$ is minimal so
\begin{equation}
m_c= \min_m\left( \arg \max_m \left( \frac{ {\bf u}(m)^{\tt T}{\bf A}{\bf u}(m) }{ {\bf u}(m)^{\tt T}{\bf u}(m)} \right) \right).
\label{eq:coreKura}
\end{equation}
The outer $\min$ function is to take into account that it is possible to have two or more $\arg \max $ values, if that is the case we choose the smallest value.

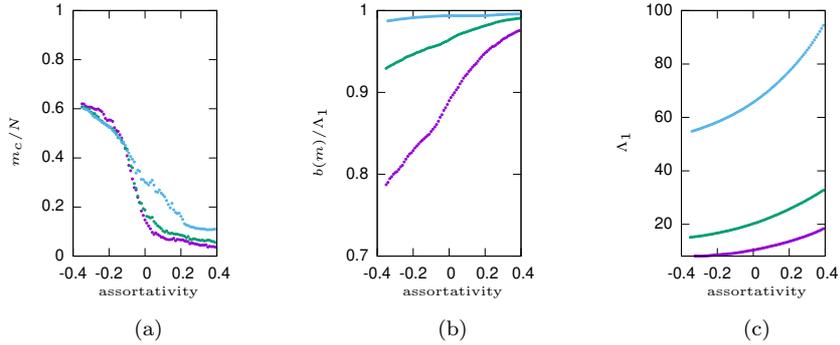
\begin{figure}
\begin{center}
\begin{tikzpicture}
\begin{scope}
\node{\includegraphics[width=3cm]{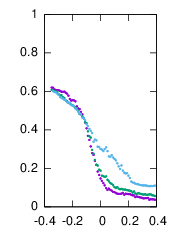}};
\draw(0.25,-2) node{\tiny \rm assortativity};
\draw(0.25,-2.5) node{\footnotesize \rm (a)};
\draw(-1.5,0) node[rotate=90]{\tiny $m_c/N$};
\end{scope}
\begin{scope}[xshift=4cm]
\node{\includegraphics[width=3cm]{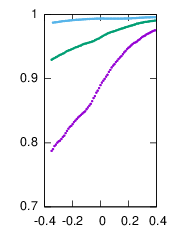}};
\draw(0.25,-2) node{\tiny \rm assortativity};
\draw(0.25,-2.5) node{\footnotesize \rm (b)};
\draw(-1.5,0) node[rotate=90]{\tiny $b(m)/\firstEigenvalue$};
\end{scope}
\begin{scope}[xshift=8cm]
\node{\includegraphics[width=3cm]{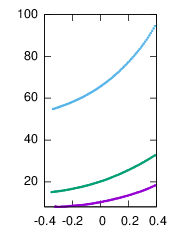}};
\draw(0.25,-2) node{\tiny \rm assortativity};
\draw(0.25,-2.5) node{\footnotesize \rm (c)};
\draw(-1.5,0) node[rotate=90]{\tiny $\firstEigenvalue$};
\end{scope}
\end{tikzpicture}
\end{center}
\caption{\label{fig:eigeAppro} (a) The core size decreases with increasing assortativity and (b) the bound improves with increasing assortativity. (c) The largest eigenvalue $\firstEigenvalue$ increases with the assortativity. The ensembles are generated using the Barabasi--Albert model with different $\degree_{\rm min}$ (purple $\degree_{\rm min}=2$, green $\degree_{\rm min}=5$ and blue $\degree_{\rm min}=20$). The number of nodes is $N=1000$. }
\end{figure}

Figure~\ref{fig:eigeAppro} 
shows an example of  the number of nodes in the core relative to total number of nodes $m_c/N$  and the ratio between the bound $b(m_c)$ and the largest eigenvalue $\firstEigenvalue$.  The figure shows three sets of networks generated using the BA model. The networks contained in a set are networks with different density of connections between the high degree nodes but with the same degree distribution and minimal degree $\degree_{\rm min}$. The change on the densities is achieved by changing the assortativity of the network. A network with positive assortativity have densely connected high degree nodes and a disassortative network have sparse connections between high degree nodes.  Given a BA network, a new network with a target assortative  is constructed using restricted randomisation~\cite{maslov2002specificity}.  

It is known that as the assortativity increases,  the eigenvalue $\firstEigenvalue$ increases~\cite{van2010influence,mondragon2017network} (Fig.~\ref{fig:eigeAppro}(c)), the new observation is that in this case,  the core decreases (Fig.~\ref{fig:eigeAppro}(a)) and the approximation to the eigenvalue improves (Fig.~\ref{fig:eigeAppro}(b)).
Also the bound $b(m)$ improves with the increase of the minimal degree.

\begin{figure}
\begin{center}
\begin{tikzpicture}
\begin{scope}
\node{\includegraphics[width=3cm]{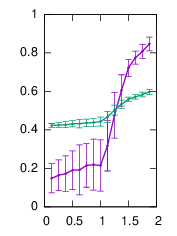}};
\draw(0.25,-2) node{\tiny $K/K_C$};
\draw(-1.5,0) node[rotate=90]{\tiny normalised order parameter};
\end{scope}
\begin{scope}[xshift=4cm]
\node{\includegraphics[width=3cm]{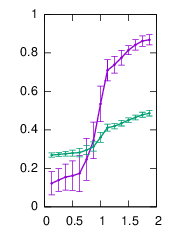}};
\draw(0.25,-2) node{\tiny $K/K_C$};
\end{scope}
\begin{scope}[xshift=8cm]
\node{\includegraphics[width=3cm]{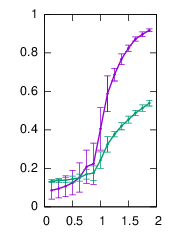}};
\draw(0.25,-2) node{\tiny $K/K_C$};
\end{scope}
\begin{scope}[yshift=-4.5cm]
\node{\includegraphics[width=3cm]{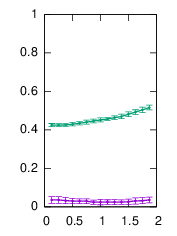}};
\draw(0.25,-2) node{\tiny $K/K_C$};
\draw(-1.5,0) node[rotate=90]{\tiny normalised order parameter};
\end{scope}
\begin{scope}[xshift=4cm,yshift=-4.5cm]
\node{\includegraphics[width=3cm]{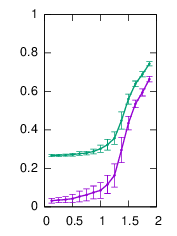}};
\draw(0.25,-2) node{\tiny $K/K_C$};
\end{scope}
\begin{scope}[xshift=8cm,yshift=-4.5cm]
\node{\includegraphics[width=3cm]{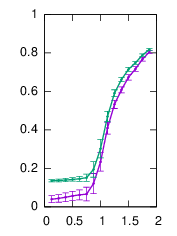}};
\draw(0.25,-2) node{\tiny $K/K_C$};
\end{scope}
\end{tikzpicture}
\end{center}
\caption{\label{fig:kuraCore} Synchronisation transition for assortative networks (top row, assortative coefficient $0.39$) and disassortative networks (bottom row, assortative coefficient $-0.40$) with different minimal degree. The behaviour of the core order parameter is shown in purple colour and the order parameter for the whole network in green. In both cases the networks were generated using the Barabasi--Albert model with $N=1000$ and minimal degree, from left to right $\degree_{\rm min}=2, 5$ and $20$. The results were generated by numerically integrating the Kuramoto Model.}
\end{figure}
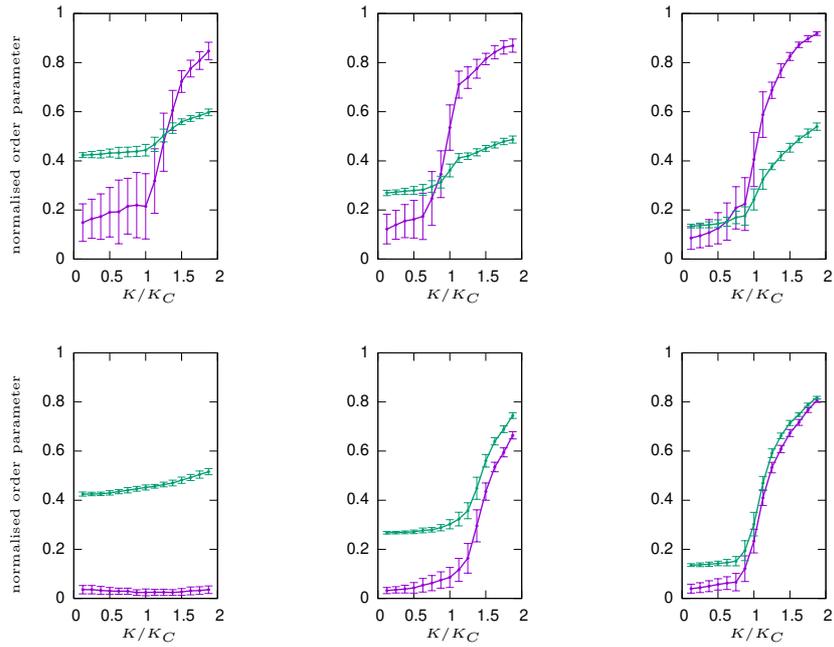

In the Kuramoto model, high degree nodes tend to synchronise more readily than other nodes~\cite{pereira2010hub}, Eq.~(\ref{eq:coreKura}) tell us how many of these high degree nodes we should consider to approximate $K_c$.  The synchronisation of the core can be measured by the local order parameter
\begin{equation}
r_c e^{i\psi_n}=\sum_{n=1}^{m_c}\sum_{m=1}^{m_c} A_{nm} \langle e^{i\theta_m}\rangle_t.
\end{equation}
Figure~\ref{fig:kuraCore} compares the beginning of synchronisation for the network and its core as a function of $K$. The network synchronisation is observed via $r=(\sum_n^N r_n)/(\sum_n^N \degree_n)$, where $r_n$ is given by Eq.~(\ref{eq:globalSynch}), and by $r_c/m_c$ for the core, so both quantities are bound in the interval $[0,1]$. The $x$--axis shows the normalised coupling $K/K_c$. For assortative network the onset of synchronisation is well approximated  by $K_c = {K_0}/{\firstEigenvalue}$ even when $\degree_{\rm min}$ is small. For disassortative networks only when the minimal degree is large the onset of synchronisation for the network and its core happens at similar values of the coupling strength.

To summarise, in the Kuramoto model where the coupling is described by a simple undirected network, it is possible to find the subset of high degree nodes that form the core of the network.   The core nodes are defined by their relevance to the phase synchronisation process and these nodes can be used to evaluate the beginning of the synchronisation for networks with positive assortative coefficient or with large minimal degree.


\begin{thebibliography}{1}

\bibitem{restrepo2005onset}
Juan~G Restrepo, Edward Ott, and Brian~R Hunt.
\newblock Onset of synchronization in large networks of coupled oscillators.
\newblock {\em Physical Review E}, 71(3):036151, 2005.

\bibitem{taubig2014matrix}
Hanjo T{\"a}ubig and Jeremias Weihmann.
\newblock Matrix power inequalities and the number of walks in graphs.
\newblock {\em Discrete Applied Mathematics}, 176:122--129, 2014.

\bibitem{maslov2002specificity}
Sergei Maslov and Kim Sneppen.
\newblock Specificity and stability in topology of protein networks.
\newblock {\em Science}, 296(5569):910--913, 2002.

\bibitem{van2010influence}
Piet Van~Mieghem, Huijuan Wang, Xin Ge, Siyu Tang, and Fernando~A Kuipers.
\newblock Influence of assortativity and degree-preserving rewiring on the
  spectra of networks.
\newblock {\em The European Physical Journal B}, 76(4):643--652, 2010.

\bibitem{mondragon2017network}
Raul~J Mondrag{\'o}n.
\newblock Network partition via a bound of the spectral radius.
\newblock {\em Journal of Complex Networks}, 5(4):513--526, 2017.

\bibitem{pereira2010hub}
Tiago Pereira.
\newblock Hub synchronization in scale-free networks.
\newblock {\em Physical Review E}, 82(3):036201, 2010.

\end{thebibliography}

\end{document}